# Signature of phase singularities in diffusive regime in one dimensional disordered lattices: Interplay and qualitative analysis

S OMNATH GHOSH[1,*]

[1]Department of Physics, Indian Institute of Technology Jodhpur, Rajasthan-342037, India
*Corresponding author: somiit@rediffmail.com

**Co-existence and interplay between mesoscopic light dynamics with singular optics in spatially random but temporally coherent disordered waveguide lattices is reported. Two CW light beams of 1.55 $\mu$m operating wavelength are launched as inputs to 1D waveguide lattices with controllable weak disorder in refractive index profile. Direct observation of phase singularities in the speckle pattern along the length is numerically demonstrated. Quantitative analysis of onset of such singular behavior and diffusive wave propagation is analyzed for the first time.**

## 1. INTRODUCTION

Lately, coupled waveguide lattices with deliberate and controllable disorder in spatial and index variations have opened up a new platform to explore and demonstrate a variety of fundamental effects [1], primarily quantum inspired photonic effects. State-of-the-art fabrication techniques have recently demonstrated many such quantum inspired physical effects [1, 2]. Moreover, existence of hidden singularities in the form of coalescence of related eigen functions (two, three or more), or simply the existence of degeneracy of eigen values are of contemporary interest in such discrete photonic structures [3–5]. These non-heritian degeneracies offer unique range of physical effects around them. Judicious choice of such point in any photonic system mimicking open quantum systems, caters solutions to many technological challenges otherwise non solvable. Unconventional photonic devices are being explored exploiting such special degeneracies of different orders. Optical isolators, asymmetric mode converters, circulators are device level implementation of such investigations. However these two domains of photonics (disordered photonic and non-hermitian singularities) have always been investigated separately for new physical insights. We realize synthesis of such mesoscopic phenomenon and the intricate network of singularities in a simplest platform should be of interest to understand new exotic phenomena qualitatively as well as quantitatively from application point of view. Hence, in this paper, we report numerical observations of the onset of disorder-induced phase singularities in speckle pattern due to interference in gain/loss assisted 1D waveguide lattices [1, 2].

Dynamics of interacting two CW light beams in the diffusive regime of light transport is chosen to explore interesting possibilities, transmission statistics and novel physical effects. Here even a weak disorder, to our surprise reveals significant physical insights from application point of view.

## 2. 1D COUPLED WAVEGUIDE LATTICE

Scalar wave propagation dynamics through waveguide lattices is governed by an equation of nonlinear Schrödinger equation like form under paraxial approximation [1, 2]:

$$i\frac{\partial A}{\partial z} + \frac{1}{2k}\left(\frac{\partial^2 A}{\partial x^2} + \frac{\partial^2 A}{\partial y^2}\right) + \frac{k}{n_0}\Delta n(x, y)A = 0, \quad (1)$$

where A(r) is the amplitude of the wave with the electric field $E(r, t) = \text{Re}[A(r)e^{i(kz-\omega t)}]$. Here, $n_0$ is the uniform background refractive index, and $\Delta n(x, y)$ represents a random deviation of the refractive index over $n_0$. Thus, refractive index profile $\Delta n(x, y)$ can be algebraically expressed in the form

$$\Delta n(x, y) = \Delta n_p(x, y)(1 + C\delta)H(x, y), \quad (2)$$

where deterministic periodic part $\Delta n_p$ of spatial period $\Lambda$, the random component $\delta$ (uniformly distributed over a specified range between 0 and 1) and dimensionless constant C, whose value governs the level of disorder in the periodic backbone structure. Moreover, along with deliberate disorder, we simultaneously incorporate an imaginary component with $\Delta n(x, y)$ as $n_{Ig}$ and $n_{IL}$ to consider gain and loss, respectively. Equation

(1) is numerically solved through the scalar beam propagation method by assuming negligible band-gap effect. We consider simultaneous launching of two Gaussian beams with plane wave-front (FWHM 10µm each) at the wavelength of 1.55µm at the unit cell located around 75$^{th}$ and 77$^{th}$ lattice position of the lattice consisting of 150 units, each of 10µm width. Thick layers of index 1.501 are separated by 5µm in a media of index 1.500. The values of $n_{Ig}$ and $n_{IL}$ are chosen to be −0.0001i and 0.0001i, respectively in low and high index layers. The overall transverse symmetry of the imaginary index distribution is maintained, however the net gain and loss is unequal. To realize lattices of certain disorder, we set different values for the disorder parameter C. The complex distribution of the scalar field is analyzed along propagation and with different lattice ensembles. The statistical nature of the singular behavior of the lattices are studied via estimation of the effective widths ($\omega_{eff}$) of the propagating beam [1], observation of network of phase singularities [3, 4] and calculation of the complexity factor defined as (where $\psi$ is complex field)

$$q^2 = <Im(\psi)^2> / <Re(\psi)^2> \quad (3)$$

## 3. LIGHT DYNAMICS: MESOSCOPIC BEHAVIOR AND SINGULARITY

To study the interplay between the presence of deliberately introduced gain/loss and wave diffusion, we have investigated the propagation of two input Gaussian beams in various disordered lattices with transverse disorder in refractive index ranging from 5% to 60%. Introduction of gain in the low index layers, results in complex coupling in the waveguide lattice. Leaky waveguides where as serve as open channels otherwise. In 1D disordered lattices with finite aspect ratio, the states are always transverse localized. However, the complex coupling chosen here has destroyed the transition to localization with increasing disorder.

**A. Correlated Speckle Formation in Diffusive Regime**

In this paper, we focus on speckle formation and onset of phase singularities other than conventional localization transition. Interaction of dual beams are so chosen that the discrete diffraction feature would be more interesting in context. Figure 1a shows the discrete speckle formation as the beams interfere in such open systems due to multiple scattering. Moreover, the intensity distributions at various lengths of the lattice in presence of 20% disorder has been depicted in Figure 1b. In the chosen lattice, as beam propagation is dynamically controlled by simultaneous disorder as well as imaginary part of the refractive index, we analyze the beam by estimating two different associated parameters. The ensemble averaged beam width (normalized with respect to initial width) has been plotted with increasing strength of transverse disorder. A clear transient diffusive regime can be found in Figure 2a. Moreover as the propagating field distribution is all complex at different length of propagation, we calculate the complexity factor over the entire range of C from 0% to 60%. Interestingly, from Figure 2b, its worth noting that the complexity saturates beyond a critical disorder level. Hence, light dynamics is certainly intriguing below the critical strength of disorder. We explore the range of lattices within this limit and investigate the interplay between wave localization and onset of phase singularity.

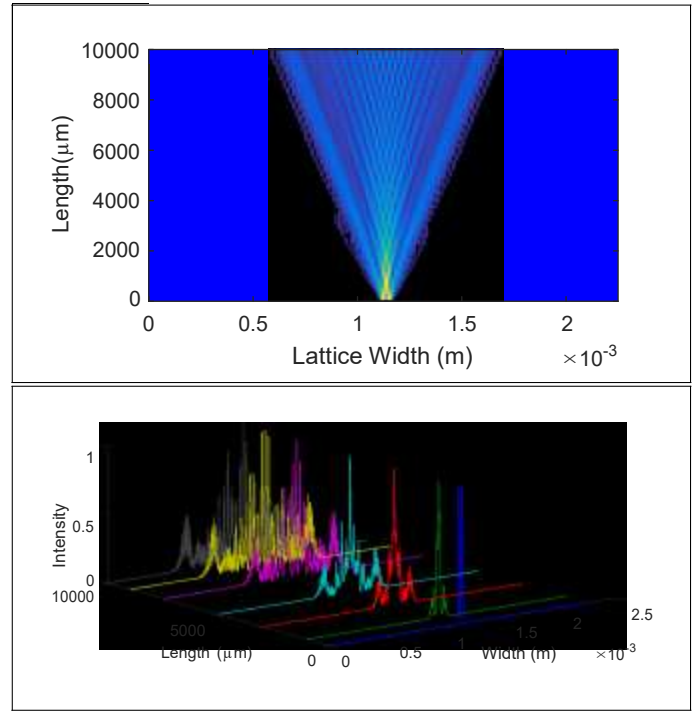

**Fig. 1(a)** Discrete diffraction through 10 mm long ordered lattice.(b) Intensity evolution of the field distributions hosting disorder (20%) induced phase singularities

**B. Network of Phase Singularities**

Purely lossy waveguide lattice (with simultaneous loss in low index layers) hosts the signature of diffusive dynamics, where as active waveguides/bulk is less interesting featuring no localized states. A careful study of the parametric dependence of the scalar beam interaction, demonstrates onset of disorder induced phase singularities with zero resultant intensity points. We analyze the complex fields along the lattice length for various configurations. However, the gain in low index layers with leaky waveguides carries signature of maximum speckle in ordered lattices when dual beam is launched. The phase plots of the beam interactions, opens up more information in the context. The phase gradient of associated field is plotted in 10% and 20% disordered lattices respectively. Figure 3 clearly demonstrates the formation of a 2D phase singularity network PSN in the lattices. From the direct comparison, the anisotropic nature of PSN distribution is evident, and the anisotropy in the PSN depends on the choice of C. As the C value is increased, the formation of phase singularities become more directional around the axis. This is first ever observation of such controlled directional movement of phase singularities in lattice geometry.

## 4. INTERPLAY AND DISTRIBUTION

Implementing transverse disorder in the form of refractive index fluctuation results in localized regime of wave propagation with increasing disorderness. Likewise, quasi-dual mode waveguides with balanced/unbalanced gain/loss encounters exceptional singularities. Synthesis of lattices with transverse disorder and simultaneous presence of gain/loss would open up a new dimension in light-state manipulation. Hence, unlike lattices with sole disorder or PT symmetric (e.g. spatially gain/loss balanced structure) lattices, our proposed platform is more interesting.

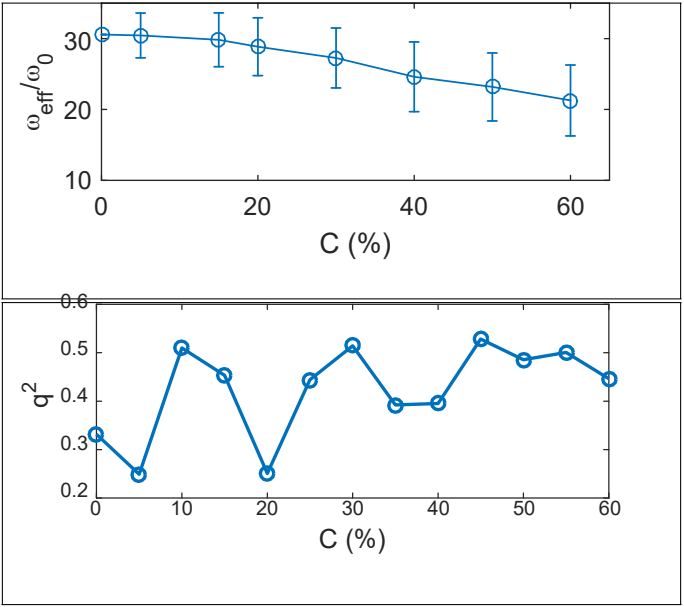

**Fig. 2.** (Color online) Variation of ensemble averaged effective beam-widths. Variation of complexity factor along the lattice length with disorder strength (C)

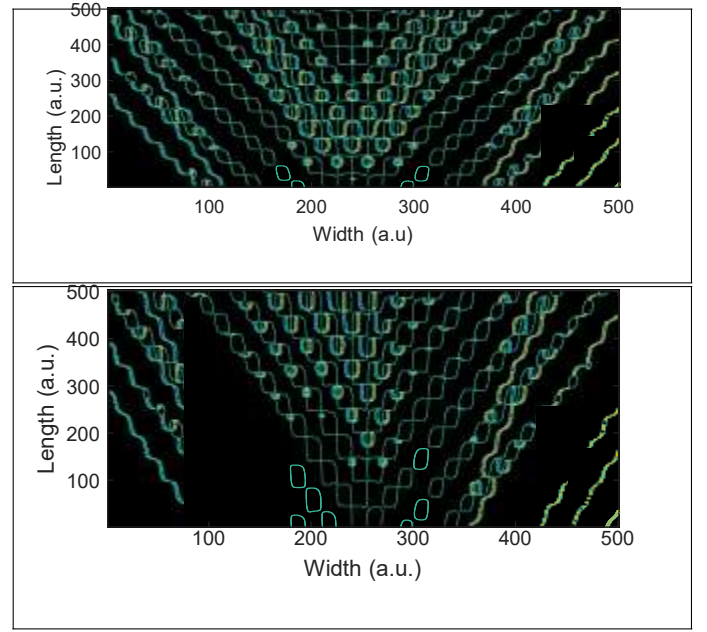

**Fig. 3.** (Color online) 2D PSN in 10% disordered (phase gradient) and (c) 20% disordered lattices (phase gradient) respectively

We deliberately exclude highly disordered lattice geometries to explore the PSN. The intriguing signature of such PSN with increasing disorder, phase front curvature and transverse wave vector of beams, have been thoroughly investigated at different operating conditions.

### A. Strength of Disorder and Transverse Wave Vector

Transverse wave vector ($k_\perp$) of the propagating beam plays the key role to dictate the speckle formation, light confinement and beam profiling in the waveguide lattices. It is inversely proportional to the effective beam width along propagation. The statistical nature of the disordered lattices with 20%, 30% and 40% disorder strengths are analyzed in terms of complexity factor and shown in Figure 4a. The narrowest beam width of (FWHM 5$\mu$m) and a wider beam of (FWHM 15$\mu$m) have been considered. The variation of q parameter attains saturation with narrower beam sizes. The q parameter fluctuates rapidly for lower values of ($k_\perp$). Hence the interplay is intriguing for wider beam sizes.

### B. Role of Phase Front

Transverse phase front has significant influence in the onset of phase singularities, and local phase derivaties at different lattice lengths.

$$A(x) = A_0 e^{-(1-iB)(x-x_0/\omega_0)^2}, \quad (4)$$

where A(x) is the amplitude of the wave associated with the Gaussian beam, B is the parameter related to phase front curvature. The 2D PSNs have discontinuities in the spatial phase gradients. Here various non zero curvature of the input beams are introduced and phase gradient of the output beam are analyzed. (B1, B2) are assigned as phase front curvature of the input beams respectively. In Figure 4b, three sets of phase front curtatures are chosen, (B1 = B2 = 0), (B1 = +1, B2 = −1) and (B1 = +1, B2 = +1). The figure evidently represents an enhancement of singular behavior along transverse direction of the lattice for non-zero curvatures. Hence dual beam configuration with curved phase front would result in denser PSN in such lattices.

### 5. CONCLUSION

Interaction between propagating light beams through waveguide lattices of different configurations and its parametric behavior has been reported. Dual mode launching condition in complex coupling environment with leaky waveguide, hosted 2D PSN in 1D photonic lattices. Sample results confirm absence of strong localization effect and saturation of complexity in the network with increasing disorder strength. Directionality along the lattice axis has been demonstrated with increasing strength of transverse disorder. Leaky waveguides with active bulk has proven to be favorable to support such singular behavior. The transverse wave-vector and finite phase front curvature control the overall light dynamics in such singular network. Our findings should be useful in non-invasive imaging, understanding of rouge wave and distinct topological phenomena.

**FUNDING INFORMATION**

Department of Science and Technology India under INSPIRE Faculty Fellow scheme (IFA-12; PH-23).

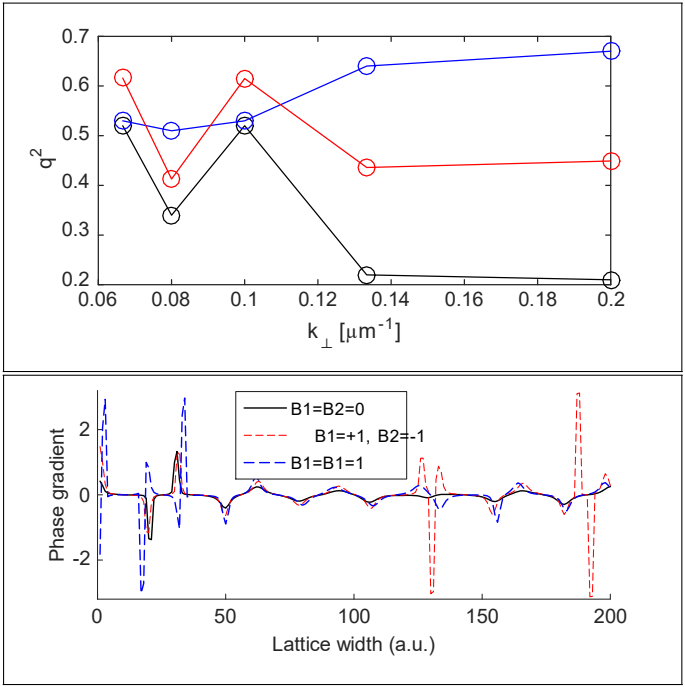

**Fig. 4.** (Color online) Variation of complexity parameters with transverse wave vector for three different disorder strengths of 20%, 30% and 40% respectively. (c) Phase gradient of the output beam with various phase front curvatures of the input beams in 20% disordered lattices.